\documentclass[aps,pra,showpacs,twocolumn]{revtex4}

\usepackage{graphicx}
\usepackage{array}

\DeclareGraphicsExtensions{.eps}

\begin{document}
\title{Dynamics of a pulsed continuous variable quantum memory}
\author{A. Dantan$^{1}$, J. Cviklinski$^{2}$, M. Pinard$^{2}$ and Ph. Grangier$^{1}$}
\affiliation{$^{1}$ Laboratoire Charles Fabry de l'Institut
d'Optique, F91403 Orsay cedex, France\\$^{2}$ Laboratoire Kastler
Brossel, UPMC, 4 place Jussieu, F75252 Paris cedex 05, France}

\date{\today}

\begin{abstract}
We study the transfer dynamics of non-classical fluctuations of
light to the ground-state collective spin components of an atomic
ensemble during a pulsed quantum memory sequence, and evaluate the
relevant physical quantities to be measured in order to characterize
such a quantum memory. We show in particular that the fluctuations
stored into the atoms are emitted in temporal modes which are always
different than those of the readout pulse, but which can
nevertheless be retrieved efficiently using a suitable temporal
mode-matching technique. We give a simple toy model - a cavity with
variable transmission - which accounts for the behavior of the
atomic quantum memory.
\end{abstract}

\maketitle

\section{Introduction}
The storage and manipulation of optical quantum states using atomic
ensembles has received considerable attention for quantum
information processing and communication \cite{lukin,duan}. Owing to
the long lifetime of ground-state atomic collective spins and the
collective coupling between the field and the atoms
\cite{julsgaard01}, atomic ensembles are good quantum registers for
quantum optical variables, and there have recently been a number of
proposals and experimental realizations of quantum state transfer
between matter and light \cite{vanderwal,matsukevitch,chou,eisaman}.
While a possible approach is to store and retrieve optical pulses
into a collective atomic excitations using the DLCZ protocol
\cite{duan}, it is also possible to map non-classical quantum
fluctuations of light to atomic ground-state spin components under
conditions of Electromagnetically Induced Transparency or Raman
resonance \cite{liu,julsgaard01,polzik}. In connection with quantum
information processing a challenging step is to store and retrieve
non-classical states. A major step in this direction has recently
been taken with the storage and retrieval of single photon pulses in
EIT \cite{singlephoton}. In the continuous variable regime, the
storage of optical coherent states has been demonstrated in atomic
vapors \cite{julsgaard04} and the mapping of squeezed or entangled
states has been studied in different configurations and different
physical systems - either in the pulsed or \textit{cw} regime
\cite{parkins,fleischhauer,kozhekin,memoire,helium,braunstein,pinard}.
For applications with continuous variable (CV) in the pulsed regime,
however, a rigorous study of the retrieval of the stored
fluctuations is needed in order to give a precise operational
definition of the quantity to be measured or utilized in quantum
information protocols.

Extending the results of Refs. \cite{memoire} to the pulsed regime,
we study in this paper the optimal conditions to transfer, store and
retrieve the fluctuations of a squeezed vacuum state to the
ground-state spins of $\Lambda$-atoms and we show that the stored
fluctuations are emitted in different \textit{temporal} mode as the
write pulse, so that optimal readout needs to be performed with a
temporally matched local oscillator.

These results, which can actually be extended to other systems than
atomic spins \cite{parkins,braunstein,pinard,helium}, provide an
operational definition of the relevant temporal modes involved in
the transfer of non-classical fluctuations of light and allow to
define quantum state transfer efficiencies for a CV quantum memory.
They also provide the link with the squeezed or entangled states
generated in pulsed experiments \cite{wenger}.

Last, we provide an equivalent toy model - a cavity with variable
transmission, which faithfully reproduces the behavior of the atomic
memory.

\section{System considered}

\begin{figure}
  \includegraphics[width=7cm]{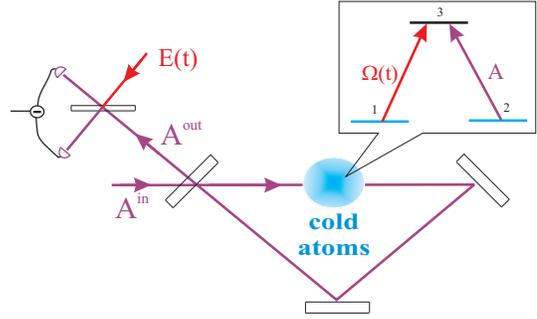}\\
  \caption{(Color online) Quantum memory set-up: field $A^{in}$ is sent into a cavity containing an ensemble of three-level
  atoms. The coupling of the intracavity field $A(t)$ with the atomic ground state variables is dynamically controlled
  by a pulsed coherent field $\Omega(t)$. The fluctuations of the output field $A^{out}$ are measured by homodyning with a local oscillator $E(t)$,
  the temporal profile of which may be adjusted so as to optimize the detection efficiency.}\label{fig:cavitelambda}
\end{figure}
Our aim is to study the dynamics of light fluctuations during write
and read sequences into a quantum memory. The physical system that
we use as a quantum memory is the collective ground-state spin
associated to two ground-state sublevels of $\Lambda$-type atoms
(Fig.~\ref{fig:cavitelambda}). The atoms interacts with a coherent
control field on one optical transition - with Rabi frequency
$\Omega(t)$, and with a field $A(t)$ possessing non-classical
fluctuations on the other transition. Since we are only interested
in this paper in the temporal aspects of the absorbed or emitted
field modes, we neglect all spatial dependence - either longitudinal
or transverse - of the fields and we assume that the atoms are
enclosed in an
optical cavity - with relatively low finesse.\\

To simplify the analysis we consider that the incoming field
$A^{in}(t)$ on the cavity is a broadband squeezed vacuum and that a
control field pulse with envelope $\Omega(t)$ (starting at
$t=-\infty$ and assumed real) is injected into the cavity. In this
case, since the vacuum field is defined with respect to the control
field, its temporal mode is precisely given by the control field
envelope. If we denote by $A(t)$ the annihilation operator
associated to the \textit{cw} multimode squeezed field, the
annihilation operator associated to the incident write pulse is
defined as \cite{grosshans}
\begin{eqnarray} \tilde{A}^{in}\equiv \frac{\int
dt\;\Omega(t)A^{in}(t)}{\sqrt{\int dt\;\Omega(t)^2}},\end{eqnarray}
the normalization being such that
$[\tilde{A}^{in},\;\tilde{A}^{in\dagger}]=1$.\\

The $N$ atoms are initially prepared in a coherent spin state:
$\langle J_z\rangle=N/2$ and $\Delta J_x^2(-\infty)=\Delta
J_y^2(-\infty)=N/4$, where $J_z=(\Pi_2-\Pi_1)/2$ is the ground-state
population difference and $J_x$, $J_y$ are the real and imaginary
parts of the ground-state coherence $J=\sum_i |2\rangle_i\langle
1|_i$. Since the squeezed field amplitude mean value is zero this
state is stationary for the mean values (assuming one neglects the
ground-state spin depolarization during the write and read pulse).
The ground-state spin quantum state is then given by the
fluctuations of $J_x$ and $J_y$, which play a role similar to that
of the field quadratures $X=A+A^{\dagger}$ and $Y=i(A^{\dagger}-A)$.
The atomic fluctuations are then only coupled to the squeezed vacuum
field when the control field is applied and it can be shown that the
vacuum field fluctuations are decoupled from those of the control
field \cite{memoire}
 \begin{eqnarray}
\delta\dot{A}&=&-\kappa\delta A+\frac{ig}{\tau}\delta
P+\sqrt{\frac{2\kappa}{\tau}}\delta
A^{in}\\\delta\dot{P}&=&-\gamma\delta P+i\Omega\delta J+igN\delta
A+F\\\delta \dot{J}&=&\;i\Omega \delta P\end{eqnarray} In order to
optimize the quantum state transfer efficiency we have assumed an
EIT-type interaction (one- and two-photon resonance) with a resonant
cavity. $\gamma$ is the optical dipole relaxation rate, $g$ the
atom-field coupling constant, $\kappa$ and $\tau$ the cavity
bandwidth and the cavity round-trip time, respectively. $F$ is a
Langevin atomic noise operator accounting for the spontaneous
emission that degrades the squeezing transfer. The squeezing
bandwidth is assumed to be broad with respect to the
atomic spectral response which will be defined later on.\\

We now assume that the interaction parameters are chosen such that
the intracavity field and the optical dipole evolve rapidly with
respect to the ground-state observables. As shown in
Ref.~\cite{memoire}, this means that the effective atomic relaxation
rate satisfy at all times,
\begin{eqnarray}\tilde{\gamma}(t)=\frac{\Omega^2(t)}{\gamma(1+2C)},\end{eqnarray} should satisfy
\begin{eqnarray}
\gamma_0\ll\tilde{\gamma}(t)\ll\gamma,\kappa\end{eqnarray} with
$\gamma_0$ the ground-state decay rate. In this case, the atomic
ground-state coherence fluctuations are linearly coupled to the
incident field fluctuations
\begin{eqnarray}\label{eq:Jx} \delta
\dot{J}_x(t)&=&-\tilde{\gamma}(t)\delta J_x(t)-\beta_E(t)\delta
X^{in}(t)+\tilde{f}_x(t),\end{eqnarray} while the outgoing field
fluctuations - $\delta A^{out}=\sqrt{\mathcal{T}}\delta A-\delta
A^{in}$ - adiabatically follow the atomic fluctuations
\begin{eqnarray}\nonumber
\delta X^{out}(t)&=&\frac{1-2C}{1+2C}\delta
X^{in}(t)-\frac{4\beta_E(t)}{N}\delta
J_x(t)\\&&\hspace{2cm}-\frac{4g}{\gamma\sqrt{\mathcal{T}}(1+2C)}F_y(t)\label{eq:Xout}\end{eqnarray}
with \begin{eqnarray}
\beta_E(t)=\frac{gN\Omega(t)}{\gamma\sqrt{\mathcal{T}}(1+2C)},
\hspace{0.5cm}\tilde{f}_x(t)=-\frac{\Omega(t)}{\gamma(1+2C)}F_y(t),\nonumber\end{eqnarray}and
$\mathcal{T}=2\kappa\tau$ is the intensity transmission of the
coupling mirror, $C=g^2N/\gamma\mathcal{T}$ is the cooperativity
parameter. Similar equations relate the orthogonal spin component
$J_y$ to the incoming and outgoing field orthogonal quadrature
$Y^{in}$ and $Y^{out}$ \cite{memoire}.

\section{Writing: atomic squeezing build-up}
Because of the linear coupling between the squeezed incident field
quadratures and the atomic spin components, the field squeezing is
transferred to the atoms during the writing phase. If we consider an
incident amplitude squeezed field with a two-time correlation
function of the form
\begin{eqnarray}\langle \delta X^{in}(t)\delta
X^{in}(t')\rangle=e^{-2r}\delta(t-t'),\end{eqnarray} it means that
the variance of the squeezed pulse is given by \begin{eqnarray}
\Delta^2 \tilde{X}^{in}=\frac{\int
dtdt'\;\Omega(t)\Omega(t')\langle\delta X^{in}(t)\delta
X^{in}(t')\rangle}{\int dt\;
\Omega^2(t)}=e^{-2r},\nonumber\end{eqnarray} consistently with our
assertion that the control field envelope defines the squeezed
vacuum pulse temporal profile.\\
Integrating (\ref{eq:Jx}) yields the normalized atomic variance -
$\Delta\bar{J}_x^2(t)\equiv \Delta J_x^2(t)/(N/4)$ - of the $J_x$
component at time $t$
\begin{eqnarray}\label{eq:varJxt}\Delta \bar{J}_x^2(t)=e^{-2a(t)}+[1-e^{-2a(t)}]
\left(\eta e^{-2r}+1-\eta\right)\end{eqnarray} with
\begin{eqnarray}
a(t)=\int_{-\infty}^{t}dt'\tilde{\gamma}(t')=\tilde{\gamma}_0\int_{-\infty}^tdt'\xi(t')\end{eqnarray}
where $\tilde{\gamma}_0=\Omega(0)^2/\gamma(1+2C)$ is the static
effective relaxation rate of Ref.~\cite{memoire}, $\xi(t)$ is the
normalized envelope of the intensity of the pulse and
$\eta=2C/(1+2C)$ is the static quantum state transfer efficiency of
\cite{memoire}. To evaluate the dynamical build-up of the atomic
squeezing, one should compare the normalized atomic noise reduction
$1-\Delta \bar{J}_x^2$ to the incident field squeezing $1-e^{-2r}$.
Taking for the control pulse a centered Gaussian profile
$\xi(t)=e^{-t^2/T^2}/\sqrt{\pi}$ of duration $T$, the quantum state
transfer efficiency at time $t$ is given by
\begin{eqnarray} \eta_w(t)\equiv \frac{1-\Delta
\bar{J}_x^2(t)}{1-e^{-2r}}=\eta [1-e^{-2a(t)}]\end{eqnarray} The
atomic squeezing increases exponentially with the integrated
intensity of the pulse $a(t)$. In steady state ($t\gg T$), the
\textit{writing} quantum efficiency is simply
\begin{eqnarray}
\eta_w=\eta(1-e^{-2a(\infty)})=\eta(1-e^{-2\tilde{\gamma}_0T})\end{eqnarray}
The efficiency $\eta_w$ approaches the \textit{cw} efficiency $\eta$
when the total pulse ``area" $a(\infty)$ is large with respect to 1
(Fig.~\ref{fig:eta}). One recovers the results of Ref.
\cite{memoire} - $\eta_w\sim\eta$ - when the pulse duration is long
with respect to the effective atomic response time
$1/\tilde{\gamma}_0$, which justifies our previous assertion that
$\eta$ is the \textit{cw} transfer efficiency. Physically, the
squeezing transfer does not depend on the profile of the write pulse
and is high when the cooperativity is large ($\eta\sim 1$), the
pulse ``area" is large enough and the incident squeezing bandwidth
is large with respect to $\tilde{\gamma}_0$.
\begin{figure}
  \includegraphics[width=7cm]{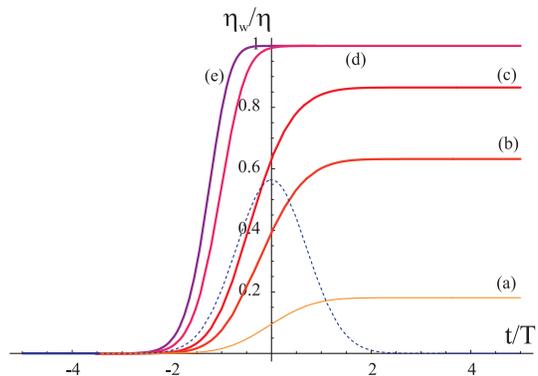}\\
  \caption{(Color online) ``Write" efficiency $\eta_w(t)$ (normalized to $\eta$) versus time $t/T$,
  when the write pulse ``area" increases [Gaussian envelopes with (a) $a(\infty)=0.1$ (b) 0.5 (c) 1 (d) 5 (e) 10].
  The dashed curve represents the write pulse envelope $\xi(t)$.}\label{fig:eta}
\end{figure}

\section{Readout}

\subsection{Outgoing field fluctuations}
\begin{figure}
  \includegraphics[width=8cm]{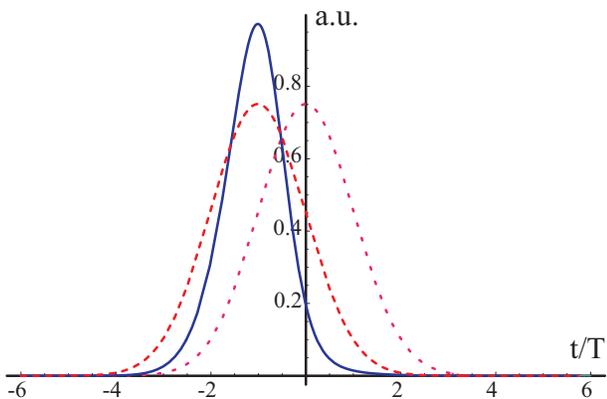}\\
  \caption{(Color online) Radiated pulse (plain), read pulse (dotted) and delayed read pulse (dashed)
  envelopes \textit{vs} time $t/T$, for $\tilde{\gamma}_0T=5$.}\label{fig:envelope}
\end{figure}

In order to readout the
atomic state after the squeezing has been stored, one can reapply -
after a variable storage time - the coherent control field, the
incident squeezed field being now turned off. The reverse process
takes place and the atoms, initially squeezed, now transfer their
squeezing to the intracavity field. This squeezing in turn reflects
in the field exiting the cavity. The evolution is still given by
(\ref{eq:Jx}-\ref{eq:Xout}), but with the initial conditions
\begin{eqnarray}\nonumber \langle \delta X^{in}(t)\delta
X^{in}(t')\rangle=\delta(t-t'),\hspace{0.4cm} \Delta
\bar{J}_x^2(-\infty)=e^{-2r}\end{eqnarray} The outgoing field
two-time correlation function can be shown to be
\begin{eqnarray}\label{eq:definitionC} \mathcal{C}(t,t')&\equiv &\langle
\delta X^{out}(t)\delta
X^{out}(t')\rangle\\
&=&\delta(t-t')+\eta
f(t)f(t')(e^{-2r}-1)\label{eq:Clecture}\end{eqnarray} with
\begin{eqnarray}\label{eq:ft} f(t)=\sqrt{2\tilde{\gamma}(t)}e^{-a(t)}\end{eqnarray}
The $\delta$-correlated term corresponds to the vacuum field
contribution that one would have without atoms or control field. The
second term carries the stored atomic squeezing with a certain
temporal profile and shows that the outgoing field will be
transitorily squeezed. The essential result is that \textit{this
temporal profile always differs from that of the readout pulse},
which means that the squeezing (and, by extension, the non-classical
fluctuations) is emitted in a different temporal mode. As can be
seen from Fig.~\ref{fig:envelope}, the field radiated by the
collective atomic dipole results in an outgoing field envelope with
a different shape than the read pulse. The maximum of emission
occurs at a time $-t_0$ different than the read pulse maximum. For a
Gaussian read pulse envelope, $f(t)\propto
e^{-[t^2/T^2+\tilde{\gamma}_0 T(1+\small{\textrm{erf}}[t/T])]/2}$
and $t_0/T=\sqrt{PL(2\tilde{\gamma}_0^2T^2/\pi)/2}$ where $PL$
denotes the product-log function. The delay between the emitted
field and the read pulse increases with the pulse area and may be
understood with the help of the toy model of Sec.~\ref{sec:toy}.

\subsection{Homodyning with the read pulse}
We now give an operational definition of the
outgoing field pulse noise measurement. We assume that its
fluctuations are measured by homodyning with a local oscillator with
a temporal profile $\mathcal{E}(t)$. The measurement operator is
then defined as
\begin{eqnarray} \tilde{X}^{out}\equiv\frac{\int dt\;
\mathcal{E}(t)X^{out}(t)}{\sqrt{\int
dt\;\mathcal{E}(t)^2}},\end{eqnarray} and its noise properties are
calculated using the correlation function (\ref{eq:definitionC})
\begin{eqnarray}\nonumber \Delta^2
\tilde{X}^{out}&=&\frac{\int dtdt'\;\mathcal{E}(t)\mathcal{E}(t')
\mathcal{C}(t,t')}{\int dt\;\mathcal{E}(t)^2}\\\nonumber
&=&1+\eta(e^{-2r}-1)\frac{\left[\int dt\;
\mathcal{E}(t)f(t)\right]^2}{\int
dt\;\mathcal{E}(t)^2}.\end{eqnarray} and the readout efficiency is
defined in a similar fashion as the write efficiency by
\begin{eqnarray}\eta_r\equiv
\frac{1-\Delta^2\tilde{X}^{out}}{1-e^{-2r}}.\end{eqnarray}

An experimentally simple and natural way to measure the field
squeezing would be to use a local oscillator (LO) that would be
(matched with) the control field read pulse and suitably delayed:
$\mathcal{E}(t)\propto\Omega(t+t_0)$. Choosing $t_0$ such that the
overlap is maximum between both pulses, the readout efficiency is
optimal - $\eta_r^*\simeq 0.96$ - when the pulse duration is such
that $\tilde{\gamma}_0T^*\simeq 2.5$ (Fig.~\ref{fig:eta}). However,
when the pulse duration is increased, the readout pulse squeezing
decreases. This imperfect efficiency can be explained by the fact
that the local oscillator in $\Omega(t+t_0)$ does not perfectly
matches the atomic emission in $f(t)$ (Fig.~\ref{fig:envelope}).
This result implies that for a practical implementation with fixed
pulse duration, the read pulse ``area" $a(\infty)$ should not be too
large in order to optimize the readout. This is in contrast with the
write sequence, for which a large ``area" pulse is preferable.
\begin{figure}
  \includegraphics[width=7.5cm]{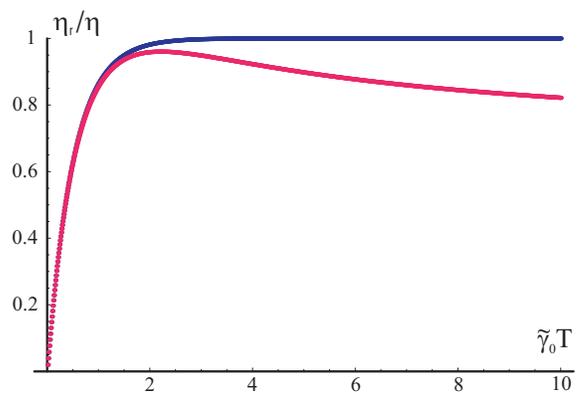}\\
  \caption{(Color online) Readout efficiency $\eta_r$ (normalized by the \textit{cw} efficiency $\eta$) versus pulse ``area" $a(\infty)=\tilde{\gamma}_0T$
  when the LO is either the retarded readout
  pulse (lower curve)
  or a pulse temporally matched with the atomic emission (upper curve).}\label{fig:eta}
\end{figure}

\subsection{Temporal matching}
It is possible however to measure the totality of the initial atomic
squeezing if one matches the LO temporal profile with that of the
field radiated by the atoms \cite{memoire,molmer}:
$\mathcal{E}(t)=\sqrt{2\tilde{\gamma}(t)}e^{-a(t)}$. In this case,
the measured field variance is then \begin{eqnarray} \Delta^2
\tilde{X}^{out}=1+\eta[1-e^{-2a(\infty)}](e^{-2r}-1)\end{eqnarray}
The shot-noise is given by $\mathcal{N}=[1-e^{-2a(\infty)}]$ and, if
one uses the same pulse to write and read, the readout efficiency is
equal to the writing efficiency
\begin{eqnarray} \eta_r=\eta[1-e^{-2a(\infty)}].\end{eqnarray}
In contrast with the previous readout method using the read pulse as
a local oscillator, the readout efficiency increases when the pulse
area is increased.

Note also that when $T$ is large with respect to the atomic response
time $\tilde{\gamma}_0^{-1}$, one recovers the \textit{cw} case of
Ref.~\cite{memoire}, in which the constant readout field is abruptly
applied at $t=0$ - the readout pulse envelope is a step function
$\Omega(t)=\Theta(t)\Omega$ - which yields a correlation function of
the form \begin{eqnarray}
\mathcal{C}(t,t')=\delta(t-t')+2\eta\tilde{\gamma}_0e^{-\tilde{\gamma}(t+t')}(e^{-2r}-1)\end{eqnarray}
The use of a local oscillator with a profile
$\mathcal{E}(t)=\Theta(t)\sqrt{2\tilde{\gamma}_0}e^{-\tilde{\gamma}_0t}$
was shown to optimize the spectrum analyzer measurement, the noise
power integrated over a time long with respect to
$\tilde{\gamma}_0^{-1}$ being the sum of a shot-noise term
$\mathcal{N}$ and a signal term $\mathcal{S}$ proportional to the
initial atomic squeezing:
\begin{eqnarray} P=\mathcal{N}+\mathcal{S}(e^{-2r}-1)\end{eqnarray}
When the integrating time is large with respect to
$\tilde{\gamma}_0^{-1}$, one has $\mathcal{S}\sim\eta\mathcal{N}$,
and thus a readout efficiency equal to $\eta$. In agreement with
this result, we can compute the outgoing pulse variance measured
with the same temporally matched LO and retrieve the same result
\begin{eqnarray}
\Delta^2\tilde{X}^{out}=1+\eta(e^{-2r}-1)\simeq
e^{-2r}\hspace{0.4cm}(\eta\sim 1)\nonumber\end{eqnarray}

\subsection{Optimal readout}

One can easily show that matching the temporal modes of the local
oscillator and the field radiated by the atoms provides the best
readout method. Indeed, for a correlation function of the form
(\ref{eq:Clecture}), the readout efficiency can be expressed as
\begin{eqnarray} \eta_r=\eta\frac{\left[\int
dt\;\mathcal{E}(t)f(t)\right]^2}{\int dt\;\mathcal{E}(t)^2}=\eta
\frac{\langle \mathcal{E}|f\rangle^2}{\langle
\mathcal{E}|\mathcal{E}\rangle}\end{eqnarray} where
$\langle.|.\rangle$ denotes the hermitian scalar product
\begin{eqnarray} \langle f|g\rangle =\int dt
f(t)^*g(t)=\int\frac{d\omega}{2\pi}f(\omega)^*g(\omega)\end{eqnarray}
In this picture, an imperfect matching between the field to be
measured and the local oscillator translates into an effective
efficiency for the homodyne detector \cite{grosshans}. Optimizing
the efficiency is clearly equivalent to maximize the overlap between
the local oscillator and the atomic emission, which yields at best
\begin{eqnarray} \eta_r^*=\eta \langle f|f\rangle\end{eqnarray} when
$|\mathcal{E}\rangle=|f\rangle$.  The best efficiency is thus
obtained when $\langle f|f\rangle=1$. With $f$ of the form
(\ref{eq:ft}), this corresponds to a constant intensity profile for
the readout pulse, such as the one of Ref.~\cite{memoire}, or pulses
of duration $T$ such that $a(\infty)\gg 1$. Let us insist on the
fact that, if it is possible to retrieve the totality of the
information on the atomic state, the fluctuations of the field
radiated during the readout sequence always have a different
temporal profile than the write pulse.

We would like to point out that this approach - which yields
convenient mathematical quantities also gives the truly physically
interesting observables for quantum information processing, as for
instance in a quantum teleportation protocol. Using the atomic
teleportation protocol of Ref.~\cite{telep} it follows from the
previous results that the optimal gain to teleport non-classical
atomic fluctuations should have a temporal profile matching the
atomic emission $f(t)$. The retrieved temporal modes takes on in
this case a clear operational definition in terms of optimizing the
teleported fluctuations.

\section{Writing and no-cloning theorem}
It is also interesting to look at what happens to the outgoing field
during the writing phase. The two-time correlation function reads
\begin{eqnarray}
\nonumber\mathcal{C}(t,t')&=&\delta(t-t')+(e^{-2r}-1)[(2\eta-1)^2\delta(t-t')\\&&+
2\eta(1-\eta)\sqrt{\tilde{\gamma}(t)\tilde{\gamma}(t')}e^{-|a(t)-a(t')|}-
\eta^2f(t)f(t')]\nonumber\end{eqnarray} When $\eta\sim 1$, the
variance measured with the matched local oscillator
$\mathcal{E}(t)=f(t)$ is simply given by
\begin{eqnarray} \nonumber\Delta^2 \tilde{X}^{out}(t)=e^{-2r}+(1-e^{-2r})\mathcal{N}(t),
\end{eqnarray} with $\mathcal{N}(t)=1-e^{-2a(t)}$, so that, at all
times, one has \cite{note}
\begin{eqnarray}
\nonumber\Delta\bar{J}^2_x(t)+\Delta^2\tilde{X}^{out}(t)=
1+e^{-2r}.\end{eqnarray} In this picture, this process clearly
appears as the quantum state transfer from one mode to another
\begin{eqnarray}\nonumber \Delta
\bar{J}_x^2(-\infty)&=&1,\hspace{0.7cm}\Delta^2\tilde{X}^{out}(-\infty)=e^{-2r}\\\nonumber
\Delta \bar{J}_x^2(+\infty)&=&e^{-2r},\hspace{0.2cm}
\Delta^2\tilde{X}^{out}(+\infty)=1.\end{eqnarray} In agreement with
the no-cloning theorem \cite{wootters}, the field squeezing
disappears while the atoms become squeezed, meaning that the initial
copy is indeed destroyed during the write sequence.

\section{Analogy with a cavity with variable
transmission}\label{sec:toy}

The atomic memory behavior can actually
be modeled in a very simple fashion by considering an empty cavity,
the transmission of which is controllable:
$\mathcal{T}(t)=2\kappa(t)\tau$ \cite{transmission}. With the same
convention as previously, the input-output relations for the field
read
\begin{eqnarray}\label{eq:dotA} \tau\delta\dot{A}(t)&=&-\frac{\mathcal{T}(t)}{2}\delta
A(t)+\sqrt{\mathcal{T}(t)}\delta A^{in}(t)\\ \delta
A^{out}(t)&=&\sqrt{\mathcal{T}(t)}\delta A(t)-\delta
A^{in}(t)\end{eqnarray} During the write sequence the incident field
is squeezed, while the intracavity field is in a vacuum state:
\begin{eqnarray}\nonumber\langle\delta
X^2(-\infty)\rangle=\frac{1}{\tau},\hspace{0.4cm}\langle \delta
X^{in}(t)\delta X^{in}(t')\rangle=\delta(t-t')e^{-2r}\end{eqnarray}
Integrating Eq.~(\ref{eq:dotA}), one gets the intracavity field
variance at time $t$ \begin{eqnarray} \langle \delta
X^2(t)\rangle=\frac{1}{\tau}\left[e^{-2r}+(1-e^{-2r})e^{-2a(t)}\right]\end{eqnarray}
with $a(t)=\int_{-\infty}^t ds\;\kappa(s)$, which has exactly the
same form as the atomic variance in Eq.~(\ref{eq:varJxt}) when
$\eta=1$ and $\tilde{\gamma}(t)$ is replaced by $\kappa(t)$. When
the incident pulse area is large - $a(\infty)\gg 1$ - the
intracavity field is perfectly squeezed when the transmission
vanishes. The squeezing is then stored into a closed cavity with
infinite lifetime \cite{lifetime}.
\begin{figure}
  \includegraphics[width=5cm]{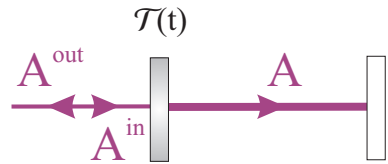}\\
  \caption{Cavity with variable transmission.}\label{fig:cavite}
\end{figure}

During the readout phase, one reopens the cavity by varying again
$\mathcal{T}(t)$. After integration with the initial conditions
\begin{eqnarray}\nonumber\langle \delta
X^2(-\infty)\rangle=\frac{e^{-2r}}{\tau},\hspace{0.4cm}\langle
\delta X^{in}(t)\delta X^{in}(t')\rangle=\delta(t-t')\end{eqnarray}
the two-time correlation function of the outgoing field takes on the
same form as in Eq.~(\ref{eq:Clecture}), with, again, $\eta=1$ and
$\tilde{\gamma}(t)$ replaced by $\kappa(t)$:
\begin{eqnarray}\mathcal{C}(t,t')=\delta(t-t')+f(t)f(t')(e^{-2r}-1)\end{eqnarray}
with $f(t)=\sqrt{2\kappa(t)}e^{-\int_{-\infty}^t ds\;\kappa(s)}$.
Within this formalism, the atomic memory considered here is clearly
equivalent to an infinite lifetime storage cavity. The
characteristics of the field emitted during readout, the transfer
efficiencies, the detection strategies are the same as for the
atomic memory case. The CV quantum memory can then be characterized
using the general methods developed previously. For instance, the
squeezing leaks out of the cavity as soon as the transmission
becomes non-zero, so that its envelope $f(t)$ is different of the
transmission profile $\mathcal{T}(t)$. The maximum of emission $t_0$
indeed depends on the transmission of the cavity at time $t_0$ as
well as the amount of squeezing that has already left the cavity.
This toy model also gives a simple interpretation for the control
field in the atomic memory scheme, which plays the same role as a
``tap" which couples or decouples the stored squeezing to the
outside.

\section{Conclusion}
We have studied the temporal mode-matching conditions for an optimal
transfer of quantum fluctuations between optical fields and an
atomic ensemble collective spin. These conditions stress the
relevant physical quantities involved in the quantum state transfer
process in \textit{cw} or in pulsed schemes. Not only do they
provide an operational meaning of the quantum states exchanged
between the fields and the atoms, but, owing to the different
temporal modes involved in the write and read sequences, they also
show how the non classical states generated in the pulsed regime
\cite{wenger} can be to measured and utilized experimentally in
continuous variable protocols – such as teleportation for instance
\cite{telep}.

We would like to point out that these results - in particular, the
simple toy model of Sec.~\ref{sec:toy} -  can easily be extended
to other storage media which can be used as continuous variable
quantum memories, such as movable mirrors or nuclear spins
\cite{pinard,helium}. In particular, a consequence for practical
implementations of continuous variable memories is that,
regardless of the storage medium, different strategies regarding
the characteristics of the read pulse can be adopted to readout
the memory.

Note also that the results derived here for squeezing are actually
valid for EPR-type entanglement or any Gaussian non-classical
fluctuations. Let us remark that the analysis developed in this
paper, in particular, he fact that quantum fluctuations are
preserved, is the analogous for the CV regime of the conservation of
the quantum character of the field in the single photon experiments
\cite{singlephoton}.

Last, in the case of an atomic memory, it would also be interesting
to look at how propagation effects will affect these results in a
scheme without cavity \cite{singlephoton,simplepassage,akamatsu}.

\begin{acknowledgments}
This work was supported by the COVAQIAL European project No.
FP6-511002.
\end{acknowledgments}


\begin{thebibliography}{99}
\bibitem{lukin} M.D. Lukin, Rev. Mod. Phys. {\bf 75}, 457 (2003).
\bibitem{duan} L.M. Duan, M.D. Lukin, J.I. Cirac, P. Zoller,
Nature (London) {\bf 414}, 413 (2001).
\bibitem{julsgaard01} B. Julsgaard \textit{et al.}, Nature (London) {\bf 413}, 400 (2001).
\bibitem{vanderwal} C.H. van der Wal \textit{et al.}, Science {\bf
301}, 196 (2003).
\bibitem{matsukevitch} D.N. Matsukevitch and A.
Kuzmich, Science {\bf 306}, 663 (2004).
\bibitem{chou} C.W. Chou
\textit{et al.}, Phys. Rev. Lett. {\bf 92}, 213601 (2004).
\bibitem{eisaman} M.D. Eisaman \textit{et al.}, Phys. Rev. Lett. {\bf 93},
233602 (2004); S.V. Polyakov \textit{et al.}, Phys. Rev. Lett. {\bf
93}, 263601 (2004).
\bibitem{liu} C. Liu \textit{et al.}, Nature (London) {\bf 409}, 490
(2001); D.F. Phillips \textit{et al.}, Phys. Rev. Lett. {\bf 86},
783 (2001); M. Bajcsy \textit{et al.}, Nature (London) {\bf 426},
633 (2004).
\bibitem{polzik} L.M. Duan, J.I. Cirac, P. Zoller, E.S. Polzik,
Phys. Rev. Lett. {\bf 85}, 5643 (2000).
\bibitem{singlephoton} T. Chaneliere \textit{et al.},
Nature (London) {\bf 438}, 833 (2005); M.D. Eisaman \textit{et
al.}, Nature (London) {438}, 837 (2005).
\bibitem{julsgaard04} B. Julsgaard \textit{et al.}, Nature (London) {\bf 432}, 482 (2004).
\bibitem{fleischhauer} M.D. Lukin
and M. Fleischhauer, Phys. Rev. Lett. {\bf 84}, 5094 (2000).
\bibitem{kozhekin} A.
Kozhekin, K. M\o lmer, E.S. Polzik, Phys. Rev. A {\bf 62}, 33809
(2001).
\bibitem{memoire} A. Dantan and M. Pinard, Phys. Rev. A {\bf 69}, 43810
(2004); A. Dantan, A. Bramati, M. Pinard, Europhys. Lett. {\bf
67}, 881 (2004).
\bibitem{parkins} A.S. Parkins and H.J. Kimble, J. Opt. B: Quantum and Semiclass. Opt. {\bf 1}, 496 (1999).
\bibitem{braunstein} J. Zhang, K.C. Peng, S.L. Braunstein, Phys. Rev. A {\bf 68}, 13808  (2003).
\bibitem{pinard} M. Pinard \textit{et al.}, Europhys. Lett. {\bf 72}, (2005).
\bibitem{helium} A. Dantan \textit{et al.}, Phys. Rev. Lett. {\bf 95}, 123002
(2005).
\bibitem{wenger} J. Wenger, R. Tualle-Brouri, P. Grangier, Opt.
Lett. {\bf 29}, 1267 (2004); J. Wenger, A. Ourjoumtsev, R.
Tualle-Brouri, P. Grangier, Eur. Phys. J. D {\bf 32}, 391 (2005).
\bibitem{grosshans} F. Grosshans and P. Grangier, Eur. Phys. J. D {\bf 14}, 119
(2001).
\bibitem{molmer} U.V. Poulsen and K. M\o lmer, Phys. Rev. Lett. {\bf 87}, 123601
(2001).
\bibitem{telep} A. Dantan, N. Treps, A. Bramati, M. Pinard, Phys. Rev.
Lett. {\bf 94}, 050502 (2005).
\bibitem{transmission} This may also correspond to a variation of the cavity
length, if one inserts a Pockels cell and a birefringent element
inside the cavity for instance.
\bibitem{note} with $\tilde{X}^{out}(t)=\int_{-\infty}^t dt'\;
E(t')X^{out}(t')/\sqrt{\int_{-\infty}^t dt' E(t')^2}$.
\bibitem{wootters} W.K. Wootters and W.H. Zurek, Nature (London)
{\bf 299}, 802 (1982).
\bibitem{lifetime} A finite lifetime could easily be included by
adding a small transmission to the second mirror for instance.
\bibitem{simplepassage} A. Dantan, A. Bramati, M. Pinard, Phys. Rev. A {\bf 71}, 43801
(2005).
\bibitem{akamatsu} D. Akamatsu, K. Akiba, M. Kozuma, Phys. Rev. Lett. {\bf 92}, 203602 (2004).
\end{thebibliography}
\end{document}